\documentclass[Physsubmission, Phys]{SciPost}

\hypersetup{
    colorlinks,
    linkcolor={red!50!black},
    citecolor={blue!50!black},
    urlcolor={blue!80!black}
}

\usepackage[bitstream-charter]{mathdesign}
\DeclareSymbolFont{usualmathcal}{OMS}{cmsy}{m}{n}
\DeclareSymbolFontAlphabet{\mathcal}{usualmathcal}

\begin{document}

\begin{center}{\Large \textbf{
Reconciling the FOPT and CIPT Predictions \\ for the Hadronic Tau Decay Rate
}}\end{center}

\begin{center}
Miguel A. Benitez-Rathgeb\textsuperscript{1},
Diogo Boito\textsuperscript{1,2} and
Andr\'e H. Hoang\textsuperscript{1,3$\star$}
Matthias Jamin\textsuperscript{1,4}
\end{center}

\begin{center}
{\bf 1} University of Vienna, Faculty of Physics, Boltzmanngasse 5, A-1090 Wien, Austria
\\
{\bf 2} Instituto de F\'isica de São Carlos, Universidade de S\~ao Paulo, CP 369, 13560-970, S\~ao Carlos, SP, Brazil
\\
{\bf 3} Erwin Schr\"odinger International Institute for Mathematical Physics, University of Vienna, Boltzmanngasse 9, A-1090 Wien, Austria
\\
{\bf 4} Department of Addictive Behaviour, Central Institute of Mental Health, Medical Faculty Mannheim, Heidelberg University, Mannheim, Germany
\\
* andre.hoang@univie.ac.at
\end{center}

\begin{center}
\today
\end{center}


\definecolor{palegray}{gray}{0.95}
\begin{center}
\colorbox{palegray}{
  \begin{minipage}{0.95\textwidth}
    \begin{center}
    {\it  16th International Workshop on Tau Lepton Physics (TAU2021),}\\
    {\it September 27 – October 1, 2021} \\
    \doi{10.21468/SciPostPhysProc.?}\\
    \end{center}
  \end{minipage}
}
\end{center}

\section*{Abstract}
{\bf
In a recent work it was suggested that the discrepancy observed in the perturbation series behavior of the $\tau$ hadronic decay rate determined in the FOPT and CIPT approaches can be explained from a different infrared sensitivity inherent to both methods, assuming that the major source of the discrepancy is the asymptotic behavior of the series related to the gluon condensate renormalon. This implies that the predictions of both methods may be reconciled in infrared subtracted perturbation theory. In this talk we explore this implication concretely in the large-$\beta_0$ approximation, where the perturbation series is known to all orders, using a renormalon-free scheme for the gluon condensate.
}

\vspace{10pt}
\noindent\rule{\textwidth}{1pt}
\tableofcontents\thispagestyle{fancy}
\noindent\rule{\textwidth}{1pt}
\vspace{10pt}

\section{Introduction}
\label{sec:intro}

Moments of the $\tau$ hadronic spectral functions obtained from LEP data~\cite{Davier:2013sfa,Ackerstaff:1998yj} can be used for precise determinations of the strong coupling $\alpha_s$. On the theoretical side, predictions of the spectral function moments involve the vacuum polarization function
$\Pi(p^2)$, where $p^2$ is the square of the invariant mass entering the quark currents.
For massless quarks $\Pi(p^2)$ is known perturbatively to 5 loops (i.e.\ ${\cal O}(\alpha_s^4)$) in full QCD~\cite{Gorishnii:1990vf,Surguladze:1990tg,Baikov:2008jh,Herzog:2017dtz}.
In the large-$\beta_0$ approximation $\Pi(p^2)$ is know to all orders~\cite{Beneke:1992ch,Broadhurst:1992si}.
Accounting only for first generation (massless) quarks, the theoretical moments can be written as~\cite{Braaten:1991qm,LeDiberder:1992jjr,Boito:2014sta,Pich:2016bdg}
\begin{equation}
\label{eq:momdef}
A_{W}(s_0) \, =\, \frac{N_c}{2} \,S_{\rm ew}\,|V_{ud}|^2 \Big[\,
\delta^{\rm tree}_{W} + \delta^{(0)}_{W}(s_0)  +
\sum_{d\geq 4}\delta^{(d)}_{W}(s_0) +\delta_W^{(\rm DVs)}(s_0)\Big] \,,
\end{equation}
where $N_c=3$, $S_{\rm ew}$ stands for electroweak corrections (which we do not consider further), $V_{ud}$ is a CKM matrix element and $s_0$ is the upper bound of the spectral function integration.
The term $\delta^{\rm tree}_{W}$ is the tree-level contribution and $\delta^{(0)}_{W}(s_0)$ stands for the higher order perturbative QCD corrections. The terms $\delta^{(d)}_{W}(s_0)$ contains power-suppressed ($\sim\!\Lambda_{\rm QCD}^{d}/s_0^{d/2}$) non-perturbative corrections obtained in the operator product expansion (OPE)~\cite{Shifman:1978bx}. The series in $d$ involves vacuum matrix elements of low-energy QCD operators of increasing dimension (conventionally called condensates) multiplied by perturbative Wilson coefficients. The leading dimension $d=4$ term is related to the gluon condensate $\langle 0|\alpha_s G^{\mu\nu}G_{\mu\nu}|0 \rangle$, which also appears in the OPE of other observables. The condensates cannot be computed from first principles and are determined together with $\alpha_s$ in simultaneous fits based on several spectral function moments.  Finally, $\delta_W^{(\rm DVs)}(s_0)$ are the duality-violation corrections that go beyond the OPE~\cite{Boito:2017cnp} and  are related to residual resonance contributions. They can be important phenomenologically, particularly for smaller choices of $s_0$  and for specific weight functions, but they will not be discussed further in this talk.

Using the 5-loop results~\cite{Gorishnii:1990vf,Surguladze:1990tg,Baikov:2008jh,Herzog:2017dtz} a number of $\alpha_s$ extractions have been carried out~\cite{Davier:2013sfa,Boito:2012cr,Boito:2014sta,Pich:2016bdg,Boito:2020xli,Ayala:2021mwc} and the world average of Ref.~\cite{ParticleDataGroup:2020ssz} now quotes a precision of about $5\%$ for $\alpha_s(m_\tau^2)$ (corresponding to an uncertainty of $1.5\%$ for $\alpha_s(m_Z^2)$). This uncertainty is dominated by the theoretical uncertainty assigned to $\delta^{(0)}_{W}(s_0)$ and involves a significant contribution related to the difference of two methods to calculate the perturbation series of $\delta^{(0)}_{W}(s_0)$~\cite{ParticleDataGroup:2020ssz}.

The QCD corrections in $\delta^{(0)}_{W}(s_0)$ and the OPE corrections in $\delta^{(d)}_{W}(s_0)$ can be obtained from the invariant mass contour integral ($x\equiv s/s_0$)
\begin{equation}
\label{eq:deltadef}
 \delta^{(0)}_{W}(s_0)  +
 \sum_{d\geq 4}\delta^{(d)}_{W}(s_0) \, =\,\frac{1}{2\pi i}\,\,
\ointctrclockwise_{{\cal C}_s}\!\! \frac{{\rm d}s}{s}\,W({\textstyle \frac{s}{s_0}})\, D(s)
\, =\,
\frac{1}{2\pi i}\,\, \ointctrclockwise_{{\cal C}_x}\!\! \frac{{\rm d}x}{x}\,W(x)\, D(x s_0)\,.
\end{equation}
where $D(s)$ is the Adler function which is related to the vacuum polarization through the relation
$\frac{1}{4\pi^2}(1+D(s)) \, \equiv \, -\,s\,\frac{{\rm d}\Pi(s)}{{\rm d} s}$. The Adler function $D(s)$ has a perturbative contribution, $\hat D(s)$, and a series of OPE corrections  ($\sim (\Lambda_{\rm QCD}^{2}/(-s))^{2,3,\ldots}$) involving the condensates already mentioned above. The weight function $W(x)$, defined in terms of the integration over the Adler function, is a polynomial in $x$ which (together with the choice of $s_0$) specifies the moment considered. The contour integration path ${\cal C}_s$ (${\cal C}_x$) starts/ends at $s=s_0\pm i 0$ ($x=1\pm i0$) and traverses the complex $s$-plane, crossing the Euclidean axis half way through, with sufficient distance from the origin such that the strong coupling stays in the perturbative regime. Through analyticity this path is related to an associated integration along the real positive $s$-axis over the experimental spectral function data with support from the pion threshold up to $s_0$~\cite{Pich:2016bdg}. Frequently a circular path with $|s|=s_0$ ($|x|=1$) is considered, but it may be deformed arbitrarily as long as it stays in the region where the strong coupling remains perturbative. The integration along such a contour path with sufficient distance from the origin ensures that the OPE series in $\delta^{(d)}_{W}(s_0)$ can be employed to parametrize non-perturbative corrections. The duality violation corrections account for non-perturbative corrections that are beyond the validity of the power expansion description of the OPE corrections and are particularly important for smaller values of $s_0$ and for specific weight function choices~\cite{Boito:2017cnp}.
For $W_\tau(x)=(1-x)^3(1+x)=1-2x+2x^3-x^4$ and $s_0=m_\tau^2$ the moment $A_{W_\tau}(m_\tau^2)$ agrees with the normalized total hadronic $\tau$ decay rate  $R_\tau=\Gamma(\tau^-\to\mbox{hadrons}\,\nu_\tau(\gamma))/\Gamma(\tau^-\to e^-\bar{\nu}_e\nu_\tau(\gamma))$.

The two widely employed methods to calculate the perturbation series $\delta^{(0)}_{W}(s_0)$ are Fixed
Order Perturbation Theory (FOPT) and Contour Improved Perturbation Theory (CIPT). The CIPT approach is based on the perturbative series for the Adler function of the form\footnote{We use the $\overline{\rm MS}$ scheme for $\alpha_s$ where the 1-loop RGE has the form $\frac{d \alpha_s(s_0)}{d\ln s_0}=-\frac{\alpha_s(s_0)^2\beta_0}{4\pi}$ with $\beta_0=11-2n_f/3$, and we furthermore define
$a(-x) \equiv \frac{\alpha_s(-s)\beta_0}{4\pi} = \frac{\alpha_s(-x s_0)\beta_0}{4\pi}$ and $a_0\equiv a(1)= \frac{\beta_0\,\alpha_s(s_0)}{4\pi} $. This implies $\frac{d a_0}{d \ln s_0}=-a_0^2$. We also take $n_f=3$ for our numerical analysis.
}
\begin{equation}
\label{eq:Adlerseries}
\hat D(s) \, =  \,
\sum_{n=1}^\infty \,
\bar c_{n} \,a^n(-x)
\, = \,
\sum_{n=1}^\infty \,
\bar c_{n} \,\,\bigg(\frac{\alpha_s(-s)\beta_0}{4\pi}\bigg)^n
\,,
\end{equation}
with real-valued coefficients $\bar c_{n}$ (which agree with those of the real-valued Euclidean Adler function for $s=-s_0$) and complex-valued powers of the strong coupling.\footnote{The common way to calculate the  $\bar c_{n}$ coefficients is to use dimensional regularization and the $\overline{\rm MS}$ scheme for $\alpha_s$ in the limit of vanishing IR cutoff.}  One carries out the contour integration over powers of the complex-valued strong coupling $\alpha_s(-s)$. The CIPT series arises from truncating the sum in Eq.\,(\ref{eq:Adlerseries}).
The FOPT approach consists of expanding the series~(\ref{eq:Adlerseries}) in powers of $\alpha_s(s_0)$, so that the complex phases appear exclusively in powers of $\ln(-s/s_0)$ within the coefficients of the power series in $\alpha_s(m_\tau^2)$. For the FOPT moments, the series arises from truncating the sum in powers of $\alpha_s(s_0)$, so that the powers of the strong coupling can be factored out of the contour integration for each series term. The CIPT approach differs from  FOPT in that it resums the powers of $\ln(-s/s_0)$ to all orders along the integration path~\cite{LeDiberder:1992jjr,Pivovarov:1991rh}.
It should also be noted that while the two expansions are related to a simple change of renormalization scheme for $\alpha_s$ at the level of the underlying Adler functions, the resulting CIPT and FOPT moment perturbation series do not have this property, since the CIPT moment series coefficients involve non-trivial integrals over powers of $\alpha_s(-s)$.

A major contribution to the theory error in $\alpha_s$ determinations from the moments $A_{W}(s_0)$ is that FOPT and CIPT calculations of  $\delta^{(0)}_{W}(s_0)$ for moments with good perturbative convergence such as the total hadronic $\tau$ decay rate\footnote{This refers to spectral function moments for which $W(x)$ does not have a quadratic term $x^2$. For these moments the gluon condensate OPE correction is strongly suppressed and the perturbation series is  much better behaved than for spectral function moments which contain such a quadratic term~\cite{Beneke:2012vb}. We elaborate more on this in Sec.~\ref{sec:numerics}.} yield systematic numerical differences that do not seem to be covered by the conventional perturbative uncertainty estimates related to renormalization scale variations. Since CIPT in general leads to smaller $\delta^{(0)}_{W}(s_0)$ than FOPT, extractions of $\alpha_s$ based on CIPT generally arrive at larger values than those based on FOPT.

\section{Motivation and basic idea of this talk}
\label{sec:essence}

In the recent work of Ref.~\cite{Hoang:2020mkw} (see also the review~\cite{Hoang:2021nlz}) it was suggested that the different behavior of the FOPT and CIPT spectral function moment series may be related to a different infrared (IR) sensitivity inherent to the FOPT and CIPT methods. It was shown that the asymptotic (i.e.\ non-convergent) character of the Adler functions's perturbation series due to infrared (IR) renormalons~\cite{Mueller:1984vh,Beneke:1998ui}  leads to a systematic and computable disparity in the behavior of the CIPT and FOPT moment series at intermediate orders where an asymptotic value is approached. This disparity, called the {\it asymptotic separation}, can be computed analytically for any weight function and for any $s_0$. Under the assumption that the bulk of the 5-loop QCD corrections to the Adler function is governed by the asymptotic behavior associated to the gluon condensate renormalon, the asymptotic separation may explain the difference between the CIPT and FOPT QCD calculations of  $\delta^{(0)}_{W}(s_0)$.

In this talk we will not repeat any of the arguments made in Refs.~\cite{Hoang:2020mkw,Hoang:2021nlz}. We start from their suggestion that the previously mentioned CIPT-FOPT disparity is of IR origin, and we explore whether this may be used to reconcile the CIPT and FOPT approaches. The final aim we have in mind is that, eventually, the discrepancy between both methods simply does not arise. Substantially more details will be presented in Ref.~\cite{conceptpaper}. Concretely, we start (i) from the assumption that the disparity between the CIPT and FOPT series is of IR origin and (ii) from the well-known fact that the OPE corrections do not only provide non-perturbative corrections but, at the same time,  should also compensate order-by-order for the asymptotically increasing behavior of the QCD corrections in the Adler function and in $\delta^{(0)}_{W}(s_0)$  ~\cite{Gross:1974jv,tHooft:1977xjm,David:1983gz,Mueller:1984vh,Beneke:1998ui}.
Property (ii) is tied to the common approach that the coefficients $\bar c_n$ in Eq.~(\ref{eq:Adlerseries}) are calculated in the limit of vanishing IR cutoff. Here we explore the idea that it should be possible to eliminate the CIPT-FOPT disparity when the IR sensitivity is removed from the perturbative calculation.

As we show in this talk, this can be achieved by using a new scheme for the condensate matrix elements which ensures that their compensating order-by-order asymptotic behavior is made explicit and can be reabsorbed into the series for the Adler function and for $\delta^{(0)}_{W}(s_0)$. We call this an IR-subtracted scheme for the condensates in contrast to the unsubtracted original scheme, which we frequently call $\overline{\rm MS}$ for brevity. Note that this does not affect the renormalization scheme for $\alpha_s$, which remains $\overline{\rm MS}$ at all times.

If constructed properly, the convergence properties of the perturbative series in the IR-subtracted condensate scheme is improved and the condensate corrections are relieved from performing duty (ii). Similar scheme changes are already well established for heavy quark masses when employing so-called short-distance mass schemes in order to remove the asymptotic behavior of the pole mass renormalon~\cite{Czarnecki:1997sz,Hoang:1998nz,Beneke:1998rk,Pineda:1998id} or the hadronization gap in soft functions for dijet event shapes~\cite{Hoang:2007vb}. In practice, care is usually only taken for the subtraction of the leading IR sensitivity, which for heavy quark masses and the dijet soft function is linear, i.e.\ ${\rm O}(\Lambda_{\rm QCD})$. For the spectral function moments the leading IR sensitivity is quartic, i.e.\ ${\rm O}(\Lambda_{\rm QCD}^4)$, and associated with the gluon condensate. Following a similar line of reasoning, we therefore only take care of the dominant quartic IR sensitivity.
Using the large-$\beta_0$ approximation, where the perturbation series is known to all orders~\cite{Beneke:1992ch,Broadhurst:1992si}, we show that this can indeed be achieved through a scheme change of the gluon condensate that is constructed
from the Euclidean Adler function for which an issue similar to the CIPT-FOPT discrepancy does not arise. We demonstrate that in this IR subtracted scheme for the gluon condensate, the discrepancy between the CIPT and FOPT calculations for $\delta^{(0)}_{W}(s_0)$ at intermediate orders is indeed much alleviated for the total hadronic $\tau$ decay. The new IR-subtracted scheme for the gluon condensate has the additional benefit that spectral function moments which behave badly in IR-sensitive $\overline{\rm MS}$ condensate schemes (and which have been argued to be unsuitable for $\alpha_s$ determinations~\cite{Beneke:2012vb}) become substantially better behaved.
We conclude with a remark on the generalization of our findings to full QCD, i.e.\ beyond the large-$\beta_0$ approximation.

\section{The Adler function and its OPE in the \texorpdfstring{large-$\beta_0$}{large-beta0} approximation}
\label{sec:largeb0}

We start by reviewing the form and structure of the perturbation series and the OPE corrections for the  Adler function in the large-$\beta_0$ approximation. The terms of its perturbative series in the $\overline{\rm MS}$ renormalization scheme for $\alpha_s$ are known to all orders and conveniently encoded in the Borel function~\cite{Broadhurst:1992si}
\begin{align}
\label{eq:AdlerBorelb0}
B[\hat D](u) & \, = \,\frac{128}{3\beta_0}\,\frac{e^{5u/3}}{2-u}\,\sum_{k=2}^\infty \, \frac{(-1)^k \,k}{[k^2-(1-u)^2]^2} \\ \nonumber
 & \, = \,\textstyle
 \frac{128}{3\beta_0}\,e^{5u/3}\Big\{
 \frac{3}{16(2-u)} +\sum\limits_{p=3}^\infty \Big[ \frac{d_2(p)}{(p-u)^2} - \frac{d_1(p)}{p-u} \Big]
 -\sum\limits_{p=-1}^{-\infty} \Big[ \frac{d_2(p)}{(u-p)^2} + \frac{d_1(p)}{u-p} \Big]
 \Big\}
\end{align}
with $d_2(p) = \frac{(-1)^p}{4(p-1)(p-2)}$ and $d_1(p) = \frac{(-1)^p(3-2p)}{4(p-1)^2(p-2)^2}$.
The coefficients $\bar c_n$ of the Adler function's perturbation series in powers of $\alpha_s(-s)$ shown in Eq.~(\ref{eq:Adlerseries}) can be recovered from the Taylor expansion
\begin{equation}
\label{eq:BTaylor}
\bigg[\, B[\hat D](u) \,\bigg]_{\rm Taylor}\, =\,
\sum\limits_{n=1}^\infty\,
\frac{u^{n-1}}{\Gamma(n)}\bar c_n
\end{equation}
and the term-by-term inverse Borel transform
\begin{equation}
\label{eq:invBorelD}
\hat D(s) = \int_0^\infty \!\! {\rm d} u \,\, \bigg[\,
B[\hat D](u)\,\bigg]_{\rm Taylor}\,e^{-\frac{u}{a(-x)}}\,,
\end{equation}
which leads to the correspondence $u^{n-1}\Leftrightarrow \Gamma(n)\,a^n$.
The coefficients of the series in powers of $\alpha_s(s_0)$ can be obtained via expansion using the relation $a(-x) = \frac{a_0}{1+a_0\ln(-x)}$ at the level of the Adler function series, or directly from Eq.~(\ref{eq:invBorelD}) rewriting it in the form
\begin{equation}
\label{eq:invBorelD2}
\hat D(s) = \int_0^\infty \!\! {\rm d} u \,\,
\bigg[\,B[\hat D](u)\, e^{-u\ln(-x)}\,\bigg]_{\rm Taylor}\,e^{-\frac{u}{a_0}}
\end{equation}
and carrying out the Taylor expansion of $B[\hat D](u)e^{-u\ln(-x)}$. We stress that here we only discuss convenient ways to write down the perturbation series of the Adler function $\hat D(s)$ for the renormalization scales $\mu^2=-s$ or $\mu^2=s_0$ in $\alpha_s$. We do not consider Eqs.~(\ref{eq:invBorelD}) or (\ref{eq:invBorelD2}) in the context of the Borel sum, i.e.\ concerning an integration over the full function in Eqs.~(\ref{eq:AdlerBorelb0}) (which has poles along the real $u$-axis). This is signified by the subscript ``Taylor".

The poles in the full Borel function $B[\hat D](u)$ at $u=p$ for positive values of $p$ , which are of IR origin, are known as IR renormalons and encode the equal-sign factorial grow of the $\bar c_n$ coefficients caused by IR momentum behavior of the loop diagrams of the vacuum polarization. A single pole term of the form $\frac{1}{p-u}$ corresponds to a contribution in the $\bar c_n$ coefficients of the form $\Gamma(n)/p^n$, and a double pole term of the form $\frac{1}{(p-u)^2}$ corresponds to a contribution in the $\bar c_n$ coefficients of the form $\Gamma(n+1)/p^{n+1}$. So, renormalons with small values of $p$ typically dominate over those with larger $p$ values, unless the normalization (or residues) of certain pole terms are strongly suppressed.\footnote{This type of suppression does not happen  in the large-$\beta_0$ approximation.} These renormalon poles encode the asymptotic character of the perturbation series of the Adler function associated to its sensitivity to small IR momenta.

\section{Infrared subtracted gluon condensate scheme}
\label{sec:IRsubtracted}

To set up the construction of the new IR-subtracted scheme for the gluon condensate, let us first talk about the Euclidean Adler function $D(-s_0)$ and its OPE.
There is a one-to-one correspondence between each IR renormalon pole, its divergent contribution to the series coefficients $\bar c_n$ and a particular term in the series for the OPE corrections, which in the large-$\beta_0$ approximation can be expressed in the form
\begin{eqnarray}
\label{eq:AdlerOPE}
\hat D^{\rm OPE}(-s_0) & = &
\frac{1}{s_0^2} \langle \bar{\cal O}_{4,0}\rangle  + \sum\limits_{p=3}^\infty \frac{1}{s_0^{p}} \Big[  \langle \bar{\cal O}_{2p,0}\rangle +
(a_0)^{-1}  \langle \bar{\cal O}_{2p,-1}\rangle  \Big]\,.
\end{eqnarray}
We note that we employ a renormalization group invariant scheme for the OPE corrections, where only the hard scale $s_0$ appears explicitly in the Wilson coefficient and the dependence on the low-energy scale is absorbed into the condensate.
Since one parametrically counts $\langle \bar{\cal O}_{2p,i}\rangle\sim \Lambda_{\rm QCD}^{2p}$, higher $p$ terms in the OPE series are formally power-suppressed as long as $s_0\gg\Lambda_{\rm QCD}^2$.
In the large-$\beta_0$ approximation
single poles at $u=p$ are related to the dimension-$2p$ condensates
$\langle\bar{\cal O}_{2p,0}\rangle$ and double poles to the dimension-$2p$ condensates $\langle\bar{\cal O}_{2p,-1}\rangle$:
\begin{equation}
\label{eq:OPEBorel}
\frac{\langle\bar{\cal O}_{2p,0}\rangle}{s_0^p} \, \Leftrightarrow \, \frac{1}{(p-u)}
\, \Leftrightarrow \, \sum_{n=1}\frac{\Gamma(n)}{p^n}\,a_0^n
 \quad {\rm and} \quad
\frac{\langle\bar{\cal O}_{2p,-1}\rangle}{s_0^p} \, \Leftrightarrow \, \frac{1}{(p-u)^2}
\, \Leftrightarrow \, \sum_{n=1}\frac{\Gamma(n+1)}{p^{n+1}}\,a_0^n \,.
\end{equation}
So the dominant gluon condensate $\langle \bar{\cal O}_{4,0}\rangle=\langle 0|\alpha_s G^{\mu\nu}G_{\mu\nu}|0 \rangle$ is associated to the simple pole at $u=2$ in Eq.~(\ref{eq:AdlerBorelb0}).
In the $\overline{\rm MS}$ scheme, the gluon condensate does not only contain a non-perturbative correction, but in addition has, order by order in $n$, to compensate for the contribution $\frac{8 e^{10/3}}{\beta_0}\frac{\Gamma(n)}{2^n}$ contained in the series coefficient $\bar c_n$ which increases factorially with $n$. This feature eventually disrupts the phenomenological use of the OPE series at high orders of perturbation theory. In most practical applications of the OPE (when the OPE corrections are still smaller than the perturbation theory terms) this is ignored with the assumption that the known QCD corrections are not at a sufficient high order so that this issue would need to be accounted for. This issue does, however, not arise using a properly designed IR-subtracted scheme for the gluon condensate.
Note that in full QCD, the OPE terms have a more complicated structure: the Wilson coefficients have higher order perturbative corrections and there are many more operators beyond the dimension-4 gluon condensate term which mix trough non-trivial anomalous dimensions. Furthermore, the renormalons in the Borel function also give rise to cuts rather than only poles.

By imposing that the compensating contribution of the gluon condensate is made explicit and that the gluon condensate correction in the IR subtracted scheme still has the form shown in Eq.~(\ref{eq:AdlerOPE}), we can now write down the relation between the original order-dependent $\overline{\rm MS}$ gluon condensate $\langle 0|\alpha_s G^{\mu\nu}G_{\mu\nu}|0 \rangle^{\overline{\rm MS},(n)}$ and a new IR-subtracted and order-independent gluon condensate $\langle 0|\alpha_s G^{\mu\nu}G_{\mu\nu}|0 \rangle(s_0)$:
\begin{equation}
\label{eq:IRsubtracted}
\langle 0|\alpha_s G^{\mu\nu}G_{\mu\nu}|0 \rangle^{\overline{\rm MS},(n)}
\, = \,
\langle 0|\alpha_s G^{\mu\nu}G_{\mu\nu}|0 \rangle(s_0)
\,- \,
s_0^2 \, \sum\limits_{i=1}^n \,\frac{8 e^{10/3}}{2^i\,\beta_0}\, \Gamma(i) \,a_0^i\,.
\end{equation}
The purpose of this IR-subtracted gluon condensate is to reshuffle the series on the RHS back into the perturbative series for the Euclidean Adler function $\hat D(-s_0)$ so that it can explicitly eliminate the effects of the gluon condensate renormalon from the original unsubtracted series. At this point, in a second step, we use the universality of the IR-subtracted gluon condensate in Eq.~(\ref{eq:IRsubtracted}), and also employ it for the OPE of the complex-valued Adler function $D(s)$. Note that here the reshuffling involves an overall factor $\frac{s_0^2}{s^2}=\frac{1}{x^2}$ since the gluon condensate correction for the (non-Euclidean) Adler function $\hat D(s)$, which has the standard form
\begin{eqnarray}
\label{eq:AdlerOPE2}
\hat D^{\rm OPE}(s) & = &
\frac{1}{s^2} \langle \bar{\cal O}_{4,0}\rangle  + \sum\limits_{p=3}^\infty \frac{1}{(-s)^{p}} \Big[  \langle \bar{\cal O}_{2p,0}\rangle +
(a_0)^{-1}  \langle \bar{\cal O}_{2p,-1}\rangle  \Big]\,,
\end{eqnarray}
involves a factor $1/s^2$ rather than the factor $1/s_0^2$ shown in Eq.~(\ref{eq:AdlerOPE}) for the Euclidean Adler function.
To ensure that the renormalon is properly cancelled, {\it it is mandatory to consistently use  exactly the same renormalization scale for the $\alpha_s$ expansion in the original unsubtracted series and the subtraction series generated by the scheme change relation~(\ref{eq:IRsubtracted}), and to truncate their sum coherently at the same order $n$}.

The IR-subtracted gluon condensate $\langle 0|\alpha_s G^{\mu\nu}G_{\mu\nu}|0 \rangle(s_0)$ we have defined at this point depends on the scale $s_0$. This is because the subtraction series depends on $s_0$, while the original $\overline{\rm MS}$ condensate $\langle 0|\alpha_s G^{\mu\nu}G_{\mu\nu}|0 \rangle^{\overline{\rm MS},(n)}$ is formally scale-independent.
This is in complete analogy to the definition of short-distance heavy quark mass schemes which all carry an explicit or implicit scale-dependence in contrast to the pole mass, which is formally scale-invariant but order-dependent and ambiguous due to the pole mass renormalon, see e.g.\ \cite{Hoang:2020iah,Beneke:2021lkq}.

Still, the $s_0$-dependence of the IR-subtracted gluon condensate defined in Eq.~(\ref{eq:IRsubtracted}) is not very convenient, and we therefore take one additional step to obtain a scale-invariant IR-subtracted gluon condensate. What we need is a closed function that has the same $s_0$-dependence as the subtraction series.
Interestingly, such a function can be obtained from the inverse Borel transform of the subtraction series when using the gluon condensate pole term from Eq.~(\ref{eq:AdlerBorelb0}):
\begin{eqnarray}
\label{eq:subtractclosed}
c_0(s_0,a_0) \,\equiv \, s_0^2\,\frac{8 e^{10/3}}{\beta_0}\,\,{\rm PV}\,\int_0^\infty \!
\frac{ {\rm d} u }{2-u}\,\,e^{-\frac{u}{a_0}}
\, = \, - s_0^2\,\frac{8 e^{10/3}}{\beta_0}\,\Gamma\bigg(0,-\frac{2}{a_0}\bigg)\,e^{-\frac{2}{a_0}}\,.
\end{eqnarray}
The integral is not unique due to the pole, and we have picked the principal value prescription, which is the common definition used for the Borel sum of the Euclidean Adler function. The prescription itself is, however, not the essential point since we can as a matter of principle add any additional constant. The relevant issue is that the derivative of $c_0(s_0,a_0)$ with respect to $s_0$ agrees with the derivative of the subtraction series summed to infinity:
\begin{eqnarray}
\label{eq:s0derivatives}
\frac{{\rm d}\,c_0(s_0,a_0)}{{\rm d}\ln(s_0) }
\,= \,
\frac{{\rm d}}{{\rm d}\ln(s_0) } \bigg[\,  s_0^2 \, \sum\limits_{i=1}^\infty \,\frac{8 e^{10/3}}{2^i\beta_0}\, \Gamma(i) \,a_0^i  \,\bigg]
\, = \,
2\,s_0^2\,e^{10/3}\,\bigg(\frac{\alpha_s(s_0)}{\pi}\bigg)\,.
\end{eqnarray}
That the derivative of the subtraction series reduces to a single term is not accidental, but a general feature of pure renormalon series in the large-$\beta_0$ approximation which are multiplied by the renormalization scale of the strong coupling raised to the power that matches the dimension of the associated condensate~\cite{Hoang:2008yj,Hoang:2009yr,Hoang:2017suc,MasterThesisRegner}. What matters for us is that the $s_0$-derivative of the (diverging) subtraction series is an extremely fast converging series (which here even reduces to a single term), so that we can use $c_0(s_0,a_0)$ to write
\begin{equation}
\label{eq:IRsubtracted2}
\langle 0|\alpha_s G^{\mu\nu}G_{\mu\nu}|0 \rangle(s_0)
\,= \,
\langle 0|\alpha_s G^{\mu\nu}G_{\mu\nu}|0 \rangle^{\rm RS} + c_0(s_0,a_0)\,,
\end{equation}
where $\langle 0|\alpha_s G^{\mu\nu}G_{\mu\nu}|0 \rangle^{\rm RS}$ defines our final IR-subtracted and scale-invariant gluon condensate. We note again that neither the exact form of the subtraction series nor the function $c_0(s_0,a_0)$ are unique.
The subtraction series merely needs to have the same asymptotic large-$n$ behavior as the one shown in Eq.~(\ref{eq:IRsubtracted}) but may have additional convergent contributions; the function  $c_0(s_0,a_0)$ has just been introduced for convenience.
So $\langle 0|\alpha_s G^{\mu\nu}G_{\mu\nu}|0 \rangle^{\rm RS}$ defines a particular scheme. Such a scheme-dependence is a general feature of renormalon subtraction schemes. Since our scheme is based on concrete and explicit expressions it can be converted to potential other schemes in a controlled and reliable manner.


We note that Eq.~(\ref{eq:IRsubtracted2}) implies the relation
\begin{equation}
\label{eq:G2relation}
\langle 0|\alpha_s G^{\mu\nu}G_{\mu\nu}|0 \rangle(s_0) - \langle 0|\alpha_s G^{\mu\nu}G_{\mu\nu}|0 \rangle(s_0^\prime)
\,= \,
c_0(s_0,a_0) - c_0(s_0^\prime,a_0^\prime)\,,
\end{equation}
with $a_0^\prime=\equiv \frac{\alpha_s(s_0^\prime)\beta_0}{4 Pi}$. This is compatible with Eq.~(\ref{eq:IRsubtracted}) since the difference of the subtraction series for $s_0$ and $s_0^\prime$ indeed sums up to $c_0(s_0,a_0) - c_0(s_0^\prime,a_0^\prime)$. To carry out the sum properly, it is (again) mandatory to expand the resulting series with a common renormalization scale in $\alpha_s$.

It is now straightforward to derive the Borel function for the perturbation series of the IR-subtracted and {\it complex-valued} Adler function $\hat D(s)$ in the form analogous to Eqs.~(\ref{eq:invBorelD}) and (\ref{eq:invBorelD2}) by parametrizing the subtraction series generated by Eq.~(\ref{eq:IRsubtracted}) back into a Borel function and taking care that the subtraction series is expanded in $\alpha_s$ having the same renormalization scale as the original unsubtracted series.
The function $c_0(s_0,a_0)$, which is conceptually part of the gluon condensate OPE correction, can -- for practical applications -- be added back to the perturbative Adler function as well. {\it We treat it like a tree-level contribution that is  not supposed to be expanded any more in powers of the strong coupling.}
The results have the form
\begin{eqnarray}
\label{eq:invBorelDs0}
\hat D_{s_0}(s) & = & \frac{c_0(s_0,a_0)}{s^2}
+ \int_0^\infty \!\! {\rm d} u \,\,
\bigg[\, B[\hat D](u)
- \frac{e^{u\ln(-x)}}{x^2}\frac{e^{10/3}}{\beta_0} \frac{8}{2-u}
\,\bigg]_{\rm Taylor}\,e^{-\frac{u}{a(-x)}}\,,
\\
\label{eq:invBorelDs02}
\hat D_{s_0}(s) & = &   \frac{c_0(s_0,a_0)}{s^2}
+\int_0^\infty \!\! {\rm d} u \,\,
\bigg[\,  B[\hat D](u)\,e^{-u\ln(-x)}
- \frac{1}{x^2}\frac{e^{10/3}}{\beta_0} \frac{8}{2-u}
\,\bigg]_{\rm Taylor}\,e^{-\frac{u}{a_0}}\,,
\end{eqnarray}
where the subscript $s_0$ indicates that the expressions refer to the perturbation series for the Adler function in our renormalon-free and scale-invariant scheme for the gluon condensate.\footnote{Even though in our renormalon-free scheme the gluon condensate is constructed to be scale-invariant, the scheme itself depends on $s_0$.} That the gluon condensate renormalon is eliminated in the subtracted series is manifest from the fact that pole at $u=2$ is cancelled inside the brackets of Eqs.~(\ref{eq:invBorelDs0}) and (\ref{eq:invBorelDs02}).

The cancellation of the gluon condensate renormalon pole entails that the major source of the disparity between the FOPT and CIPT series for the total hadronic $\tau$ decay rate is eliminated, and that both methods will provide compatible descriptions. This is indeed happening, as we show below.
As already mentioned, a valuable benefit of our IR-subtracted and scale-invariant gluon condensate scheme is that now also spectral function moments with weight functions containing a quadratic $x^2$ term can be expected to be perturbatively well-behaved (and therefore more useful for phenomenological analyses). We also mention that our scheme has the additional nice feature that the subtracted perturbation series for the spectral function moments are numerically close to the original unsubtracted series at intermediate orders.

\section{Brief numerical analysis}
\label{sec:numerics}

We now analyse briefly the behavior of the spectral function moment perturbation series for the kinematic weight function $W_\tau(x)=(1-x)^3(1+x)=1-2x+2x^3-x^4$ (which provides predictions for the total hadronic $\tau$ decay rate for $s_0=m_\tau^2$) and for $W(x)=(1-x)^3=1-3x+3x^2-x^3$, both of which are doubly-pinched (i.e.\ vanish with cubic strength) at $x=1$. In the following we compare the behavior of the original series and of the IR-subtracted series which is free of the gluon condensate renormalon. We determine the CIPT moment series from the perturbation series for $\hat D_{s_0}(s)$ in Eq.~(\ref{eq:invBorelDs0}) and FOPT moment series from 
Eq.~(\ref{eq:invBorelDs02}).

\begin{figure}
	\centering
	\includegraphics[width= 0.95\textwidth]{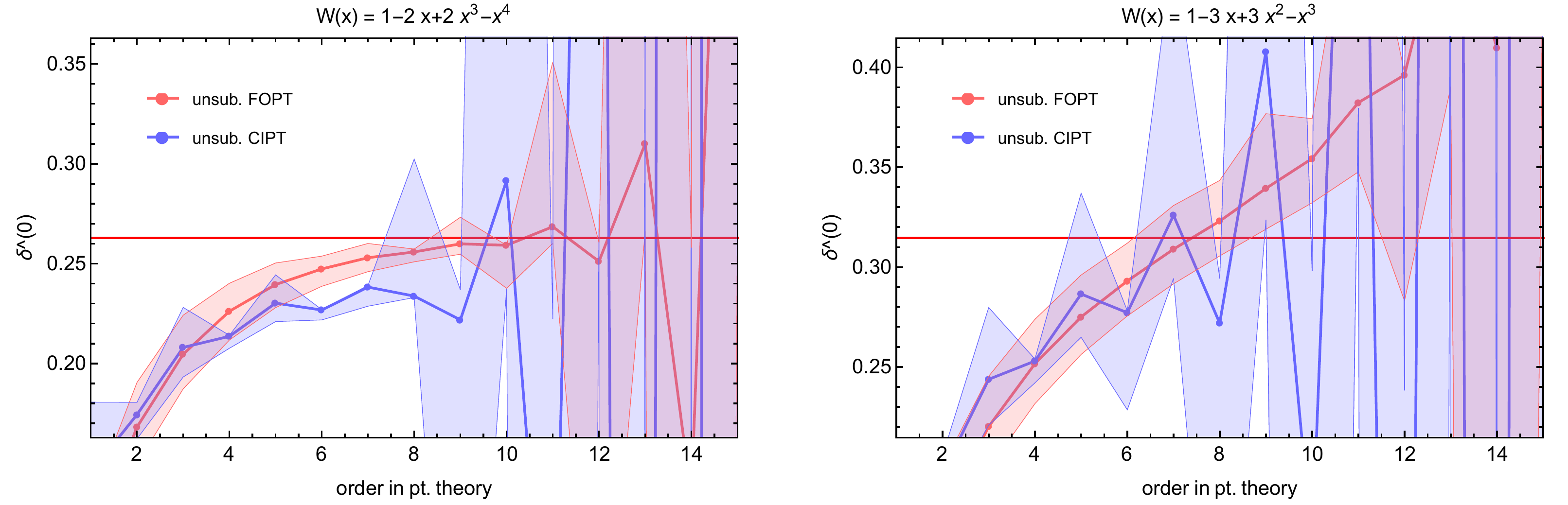}\\
	\includegraphics[width= 0.95 \textwidth]{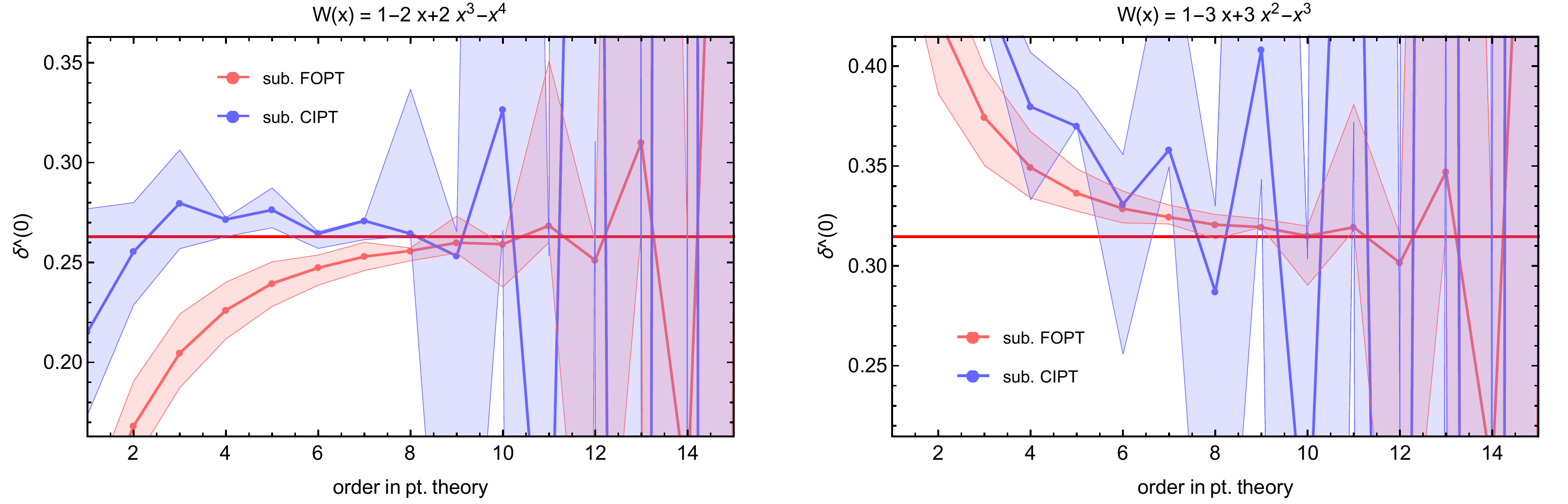}
	\caption{\label{fig:moments} \small
		FOPT and CIPT spectral function moment series in the large-$\beta_0$ approximation for the kinematic weight function $W_\tau(x)$ (left panels) and for $W(x)=(1-x)^3$ (right panels). The upper panels show the results when the $\overline{\rm MS}$ scheme is used for all condensates, the lower panels are for our renormalon-free gluon condensate scheme. The red line is the FOPT Borel sum (defined with PV prescription) which serves mainly as a line of orientation.}
\end{figure}

The kinematic weight function $W_\tau(x)$ does not contain a quadratic $x^2$ term. As a consequence, when carrying out the contour integral~(\ref{eq:deltadef}) in the FOPT approach, the subtraction series and the term $c_0(s_0,a_0)/s^2$ generated through our IR-subtracted gluon condensate scheme vanish in the same way as the gluon condensate OPE correction vanishes. These facts illustrate that all the effects related to the gluon condensate renormalon are eliminated for the kinematic weight function. This is the reason why weight functions without a quadratic term lead to more stable predictions~\cite{Beneke:2012vb,Boito:2020hvu} (see also footnote~3). As a consequence, the FOPT perturbation series in the subtracted gluon condensate scheme is the same as in the original $\overline{\rm  MS}$ gluon condensate scheme. For the CIPT method the
contribution from $c_0(s_0,a_0)/s^2$ and the gluon condensate correction are eliminated as well, but the subtraction series terms do not vanish at any order. This by itself is not a problem, because it is expected to happen from the way the subtraction scheme has been constructed.
The important question is how the subtraction series behaves at higher orders. If the CIPT moment series had the same IR sensitivity as the FOPT moment series, the subtraction series would converge to zero.

Let us have a look at the numerical results shown in Fig.~\ref{fig:moments}, where the results for the FOPT (red) and CIPT (blue) kinematic moment series for $s_0=m_\tau^2$ are shown in the original unsubtracted $\overline{\rm MS}$ gluon condensate scheme (upper left panel) and in the renormalon-free IR-subtracted scheme (lower left panel) up to order 15. The dots (and connecting lines) refer to the $\alpha_s$ renormalization scale $\mu^2=m_\tau^2$ for FOPT and $\mu=-s$  for CIPT and the corresponding bands refer to the scale variations of the form  $\mu^2=\xi m_\tau^2$ and  $\mu=-\xi s$, respectively, for $\frac{1}{2}\le \xi\le 2$. All results have been obtained with $\alpha_s(m_\tau^2)=0.34$. For the original $\overline{\rm MS}$ scheme series (upper left panel) we clearly see the well-known CIPT-FOPT discrepancy: while both series are compatible at very low orders, at orders 5 to 7 there is a systematic disparity of the CIPT series being below the FOPT series. (At order 8 and beyond the sign-alternating effects of UV renormalons begin dominating the series and scale variations increase strongly, so that considering only IR renormalon effects in discussing the results is not sufficient~\cite{Beneke:2008ad}.) In the IR-subtracted scheme (lower left panel) we see, as already mentioned above, that the results for the FOPT series is unchanged. On the other hand, the behavior of the CIPT series is significantly modified. It is now located above the FOPT results and does not anymore undershoot FOPT. Its description is now more consistent with the FOPT series within the scale variation uncertainties. Furthermore, the two series approach consistent values before the divergent behavior sets in. Interestingly, the CIPT series is much faster converging than the FOPT series.

On the conceptual side, we see that the subtraction series for CIPT clearly does not converge to zero.
Rather, the results indicate that the subtraction series suppresses the discrepancy the unsubtracted CIPT series had with respect to the FOPT series. This observation corroborates the statement made in Refs.~\cite{Hoang:2020mkw,Hoang:2021nlz} that the CIPT and FOPT perturbation series have different IR sensitivity. Once a renormalon subtracted scheme for the gluon condensate is employed, the issue appears to be alleviated. Overall,  {\it the results confirm that the CIPT-FOPT discrepancy in the original unsubtracted scheme is of IR origin, and we also find that using a renormalon-free gluon condensate scheme visibly leads to CIPT and FOPT predictions that are consistent}.

We conclude the analysis by also having a look at the CIPT and FOPT perturbation series for a spectral function moment based on the weight function $W(x)=(1-x)^3=1-3x+3x^2-x^3$ which contains a quadratic term. For this moment the subtraction series does neither vanish in CIPT nor in FOPT.
In the unsubtracted $\overline{\rm MS}$ scheme for the gluon condensate (upper right panel) we observe that the CIPT and FOPT series terms, both, do not at all decrease with order. This illustrates the bad perturbative behavior of spectral function moments with weight functions that contain a quadratic term and why such moments are not suitable for phenomenological analyses. However, using our renormalon-free scheme for the gluon condensate (lower right panel) the perturbative behavior is visibly improved at least for the FOPT series. This observation is highly encouraging. It suggests that spectral function moments with weight function containing a quadratic term may become eligible for precision phenomenological analyses in renormalon-free condensate schemes. For the CIPT series the scale uncertainties increase so strongly with order that no conclusion can be drawn. This is related to an enhancement of the sign-alternating UV renormalon asymptotics that takes place only for the CIPT approach~\cite{Beneke:2008ad}.

\section{Conclusion}
\label{sec:conclusion}

In this talk we have shown in the large-$\beta_0$ approximation, where all-order results exist, that the discrepancy between the CIPT and FOPT perturbation series for the total hadronic $\tau$ decay rate, which arises in the standard $\overline{\rm MS}$ scheme for the OPE matrix elements, does not take place when a renormalon-free IR-subtracted scheme is used for the gluon condensate. We have explicitly constructed such a renormalon-free scheme. Using an IR-subtracted scheme for the gluon condensate is sufficient in large-$\beta_0$ since the asymptotic series behavior of the Adler function associated to the gluon condensate dominates the perturbative coefficients already at intermediate orders.
Our results are consistent with the suggestion made in Refs.~\cite{Hoang:2020mkw,Hoang:2021nlz} that the discrepancy between the CIPT and FOPT perturbation series in the standard $\overline{\rm MS}$ scheme for the OPE matrix elements is of IR origin, and that different OPE corrections should be employed for both methods. Our results also imply that both issues can be remedied when renormalon-free schemes for the condensates are used.  We found that using a renormalon-free scheme for the gluon condensate significantly improves the perturbative behavior of spectral function moments where the gluon condensate OPE correction is not suppressed, such that they may become useful for high-precision phenomenological analyses.

We conclude with a remark in which way our findings can be generalized to full QCD. In full QCD the actual analytical and numerical results are more involved and will of course differ. But the construction of the IR-subtracted and scale-invariant gluon condensate is straightforward using the known renormalon calculus, and the conceptual aspects are unchanged.
However, a limitation arises since we do not have precise knowledge on the normalization of the gluon condensate renormalon singularity in the full QCD analogue of the Borel function in Eq.~(\ref{eq:AdlerBorelb0}). Since it is not possible to determine the norm of the singular and non-analytic structure related to the gluon condensate renormalon from first principles in full QCD (the normalization is of non-perturbative origin) an IR-safe scheme for the gluon condensate can only be constructed with additional assumptions. We refer the reader to Ref.~\cite{conceptpaper} for details.

\section*{Acknowledgements}
DB and MJ thank the University of Vienna for hospitality.  DB's work was supported in part by  Coordenação de Aperfeiçoamento de Pessoal de Nível Superior – Brasil (CAPES) – Finance Code 001. We thank Christoph Regner for discussions.

\bibliographystyle{SciPost_bibstyle}
\bibliography{sources.bib}

\nolinenumbers

\end{document}